 \documentclass[keywords,a4paper,11pt]{article}
\usepackage{amsmath}
\usepackage{amssymb}
\usepackage{amsfonts}
\usepackage{latexsym}
\usepackage{epsfig}
\usepackage{float}
\makeatletter
\renewcommand\@biblabel[1]{#1.}
\makeatother
\begin{document}

\title{Localization and characterization of simple defects in finite-size photonic crystals}


\author{Jean-Philippe Groby\thanks{$^1$Centre de Math\'ematiques Appliqu\'ees, UMR7641 CNRS/\'Ecole Polytechnique Route de Saclay, Palaiseau,  F-91128}, Dominique Lesselier\thanks{D\'epartement de Recherche en
\'Electromagn\'etisme - Laboratoire des Signaux et Syst\`{e}mes, UMR8506 
(CNRS-Sup\'elec-Univ Paris-Sud) Gif-sur-Yvette, F-91192}}

\maketitle

\begin{abstract}
Structured materials like photonic crystals require for optimal use a high precision both on position and optical characteristics of the components  which they are made of. Here, we present a simple tomographic algorithm, based on a specific Green's function together with a first-order Born approximation, which enables us to localize and characterize identical defects in finite-size photonic crystals. This algorithm is proposed as a first step to the monitoring of such materials. Illustrative numerical results show in particular some possibility of focalization beyond the Rayleigh criterion.
\end{abstract}

\section{Introduction}
Photonic crystals (PCs) \cite{Joannopoulos,Yablonovitch} are periodic, dielectric or metallic structures, which possess a variety of band dispersions and band gaps.  They are found in antennas \cite{Enoch,Brown}, waveguides \cite{lalanne}, negative refractive index materials \cite{foteinopoulou} to quote a few usages. As is well-known, such structures  require for optimal use that both position and optical properties of materials they are consisting of be very precise. Though the density of states is zero within the photonic band gap, by perturbing a single lattice site, a single mode or a set of closely spaced modes which have frequencies within this gap is permitted, e.g., a single column can be removed from the crystals or replaced with another the size, shape, or dielectric constant of which is different from the original. Properties \cite{Ho,Asatryan, Leung} and modeling \cite{Tayeb,Botten} of PCs have been extensively studied, though mostly for 2-D configurations, these last two decades. 

The present investigation is intended to be a first step towards the health monitoring of such structures. A low-complexity tomographic algorithm that enables us to localize and characterize simple defects consisting in either an absence of circular cylinders or a modification of the optical properties of such cylinders in finite-sized 2-D photonic crystals. This is done from a low-frequency excitation outside the usual frequency range of the band gap. Both the specific Green's function and the response to the low-frequency excitation of the original, intact structure (and of the damaged structure) can be calculated by means of the so-called multipole method \cite{Asatryan,Felbacq}, and they can be used to solve the inverse problem via a first-order Born approximation. The introduction of a specific Green's function directly exhibits the discrepancies between the initial configuration and the configuration with defects \cite{groby}.

In the background of this investigation, recent studies have also shown that the utilization of a structured embedding medium, like PCs, could lead to focusing and resolution of a tomographic inversion algorithm beyond the Rayleigh criterion \cite{ozbay,Sentenac}. The possibility to encounter such a phenomenon is investigated in addition herein.

\section{Field (Data) and specific Green's function.}

Let us consider a structure made of $N$ parallel circular cylinders $\mathcal{C}^j$, identified by superscript $j\in N$, of radius $R^j$ and of optical index $\eta^j$, located at $\mathbf{r}^j=(r^j,\theta^j)$ in the global polar coordinate system in the cross-sectional plane (this is a 2-D scattering configuration).

As indicated in the above, both fields and a specific Green's function (the latter being the field solution of the problem when the structure is excited by a given line source) are calculated by means of the multipole method \cite{Asatryan,Felbacq}. Key to this approach are the local field expansions or multipole expansions in the vicinity of each cylinder in the polar coordinate system linked to that cylinder (which are derived from the application of the Graf's addition theorem \cite{Chew}):

\begin{equation}
E^{e}(\mathbf{r_l})=\sum_{m=-\infty}^{\infty}\left[ B_m^{l}
H_m^{(1)}\left(k r_l\right)
+A_m^l J_m\left(k r_l\right)\right]e^{\mbox{i}m\theta_l}
\label{s1e1}
\end{equation}

\noindent wherein $H_m^{(1)}$ is the first-kind Hankel function of order $m$, $J_m$ is the Bessel function of order $m$, $B_m^{l}$ are the coefficients of the scattered field by the $l$-th cylinder, $A_m^l$ are those of the incident field impinging upon the $l$-th cylinder, and $\mathbf{r_l}=(r_l,\theta_l)$ are the coordinates of a point P in the polar coordinate system linked to this $l$-th cylinder. The local incident field on the $l$-th cylinder is generated by the actual incident field $E^{inc}$ as well as by the fields that are scattered by all other cylinders $j$, $j \neq l$. Their coefficients also take the form

\begin{equation}
A_m^l=K_m^l+\sum_{j=1,j\neq l}^{N} \sum_{p=-\infty}^{\infty} S_{mp}^{lj} B_p^j
\label{s1e2}
\end{equation}

\noindent wherein $K_m^l$ are the coefficients of the actual incident field (either a planar incident wave in the cross-sectional plane, $ K_m^l=(-\mbox{i})^m \exp \left({-\mbox{i} k r^l \cos\left(\theta^{inc}-\theta^l \right)-\mbox{i}m\theta^{inc}}\right)$, or a cylindrical wave generated by an exterior line source set parallel to the axis of the cylinder) and $S_{mp}^{lj}=H_{m-p}^{(1)}\left(k r_l^j\right)
e^{\mbox{i}(p-m)\theta_l^j}$ are translation terms, $(r_l^j,\theta_l^j)$ being the coordinates of the $j$-th cylinder in the polar coordinate system associated to the $l$-th cylinder.

Coefficients $A_m^l$ and $B_m^l$ are related  together via the continuity of the tangential components of the electric and magnetic fields to be imposed at the cylinder boundaries. To derive these relationships, the interior field expansion within the cylinder $l$ is used as

\begin{equation}
E^{l}(\mathbf{r_l})=\!\!\!\sum_{m=-\infty}^{\infty}\left[ Q_m^{l}
H_m^{(1)}\left(k \eta^l r_l\right)
+C_m^l J_m\left(k\eta^l r_l\right)\right]e^{\mbox{i}m\theta_l}
\label{s1e3}
\end{equation}

\noindent wherein $C_m^{l}$ are the coefficients of the scattered field inside the $l$ cylinder and $ Q_m^{l}=\frac{\mathbf{i}}{4}\chi^l J_m(k \eta^l r_l^s)e^{\mathbf{i}m\theta_l^s}$ are the coefficients of a field generated by a line source located at $(r_l^s,\theta_l^s)$ inside the cylinder $l$ in the polar coordinate system associated to it. The presence of the interior source is indicated by the term $\chi^l$ valued to  $1$ when the source is present and $0$ otherwise.

The continuity conditions at the boundaries are most conveniently expressed in terms of cylindrical harmonic reflection and transmission coefficients \cite{Asatryan} as

\begin{equation}
\begin{array}{l}
B_m^l=R_m^lA_m^l+T_m^lQ_m^l\\
C_m^l=T_m^{'l}A_m^l+R_m^{'l}Q_m^l.\\
\end{array}
\label{s1e4}
\end{equation}

In the above, vectors $\mathbf{B}=[B_m^l]$, $\mathbf{K}=[K_m^l]$ and $\mathbf{Q}=[Q_m^l]$, as well as matrices  $\mathbf{S}=[S_{mp}^{jl}]$, $\mathbf{R}=\mbox{diag} R_m^l$ (with identical definition applying to the other reflection and transmission matrices and the other coefficients), are introduced in order to deduce from (\ref{s1e2}) and (\ref{s1e4}) the system of linear equations in the source coefficients  $\mathbf{B}$ as

\begin{equation}
\left(\mathbf{I}-\mathbf{R}\mathbf{S}\right)\mathbf{B}=\mathbf{R}\mathbf{K}+\mathbf{T}\mathbf{Q}.
\label{s1e5}
\end{equation}

Upon solving the above linear system both in the interior (using the second equation of (\ref{s1e4})) and in the exterior, the specific Green's function is made available.

\section{A simple tomographic algorithm.}

\begin{figure}[H]
  \centerline{\includegraphics[width=7.0cm]{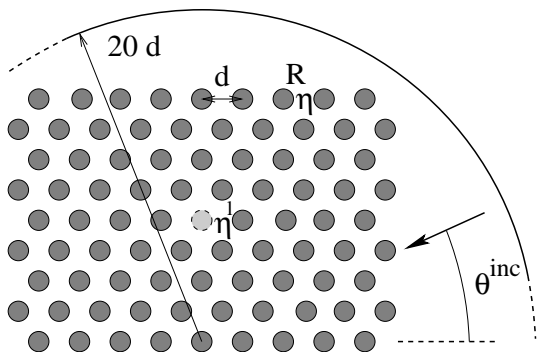}}
  \caption{Finite crystals of $N=85$ parallel circular cylinders with a single defect. The circle around the structure is the one used for the field computation.}
\label{conf}
  \end{figure}

From now on, one is considering a finite-size crystal (FSC) with hexagonal symmetry as is sketched in Fig. \ref{conf}. It is made of $N$ ($N$ will be chosen as $85$ in the numerical examples) circular cylinders of same radius $R$ and same optical index $\eta$. The cylinders $\mathcal{C}^j$, $j\in[1,N]$ are ordered such that $\mathcal{C}^1$ is located at the bottom left corner and $\mathcal{C}^N$ at the top right corner. The distance between the centers of the closest cylinders is denoted by $d$. The electric field is calculated on a circle of radius $r=20d$ (this is a rather arbitrary value, what matters is that one stays fully outside the crystal). Fields are time-harmonic,  with wavelength in free space as $\lambda$, and wavenumber $k=2\pi/\lambda$. A defect is obtained either by removing the $l$-th cylinder, or by modifying its optical index, de facto creating a novel configuration (denoted as $l$-th DFSC).

Tomographic algorithms are usually derived from an integral formulation of the solution based on the so-called background (free-field) Green's function. Use of a specific Green's function, i.e., the solution (for a given line source) of a problem close to the one at hand reduces the kernel of the integral. Here, since the structure of materials which one is interested in is designed to exhibit specific properties, it is consistent to consider the Green's function of the configuration associated with this structure (i.e., in the absence of defects) whenever intending to carry out the health monitoring.

\subsection{Localization and characterization of a single defect}

One introduces $g(\mathbf{r},\mathbf{r_l^s})$ and $E(\mathbf{r})$ as the specific Green's function and the total electric field  calculated for the FSC, and $E^{l}(\mathbf{r})$ the total electric field calculated for the $l$-th DFSC. These fields are related by the well-known Fredholm integral equation of first kind $ E^{l}(\mathbf{r})-E(\mathbf{r})=\int_{\mathcal{C}^l} g(\mathbf{r},\mathbf{r_l^s}) k^2\left(\left(\eta^l\right)^2-\eta^2\right)E^{l}(\mathbf{r_l^s}) d\!S$.

Then, a first-order Born approximation, whose accuracy is related to the fact that the field at the location of the defect should be weakly modified by the defect, is employed. Any mode associated to the defect, possibly leading to entrapment of the field, should be excited. Complementarily, one applies a low-frequency approximation, which requires that the radius of each cylinder is small enough with respect to the source wavelength. Furthermore, one works outside the band gap. Under these conditions the Green's function and the field $E^l(\mathbf{r_l^s})$ read as

\begin{equation}
\begin{array}{l}
\displaystyle g(\mathbf{r},\mathbf{r_l^s})=\frac{\mbox{i}T_{0}}{4}J_{0}\left(k\eta r_l^s \right)\sum_{j=1}^{N}b_{0}^{j}\left(\chi^l \right)H_{0}^{(1)}\left(k|\mathbf{r}-\mathbf{r^j}| \right)\\
\displaystyle E^l(\mathbf{r_l^s})=\mathcal{C}_0^l J_{0}\left(k\eta r_l^s \right)=T_{0}^{'} \mathcal{A}_0^l J_{0}\left(k\eta r_l^s \right)
\end{array}
\label{s2e1}
\end{equation}

In Eq.(\ref{s2e1}) one has introduced $b_{0}^{j}$, such that $ \frac{\mbox{i}}{4}T_{0}J_{0}\left(k\eta r_l^s \right)b_{0}^{j}=B_{0}^j$ is satisfying the linear set of equations $\left(\mathbf{I}-\mathbf{R}\mathbf{S} \right)\mathbf{b}=\boldsymbol{\chi}$, with vector $\boldsymbol{\chi}=[\chi^l]$. Coefficients of the locally impinging field $\mathcal{A}_0^l$ can be calculated via the solution of Eq.(\ref{s1e2}) with $\mathcal{B}_{0}^j$ satisfying $\left(\mathbf{I}-\mathbf{R}\mathbf{S} \right)\boldsymbol{\mathcal{B}}=\mathbf{R}\boldsymbol{\mathcal{K}}$, $ \mathcal{K}_0^j= e^{-\mbox{i} k r^j \cos\left(\theta^{inc}-\theta^j \right)}$.

By making use of relation \cite{Gradsteyn} $ \int x \left(J_0(\alpha x) \right)^2 d\!x=\frac{x^2}{2}\left\{\left(J_0(\alpha x) \right)^2+\left(J_1(\alpha x) \right)^2 \right\}$, the first-order Born approximation of the integral equation is

\begin{equation}
\displaystyle E^{l}(\mathbf{r})-E(\mathbf{r})\approx \mathcal{D}\zeta^l\mathcal{A}_0^l \sum_{j=1}^N b_{0}^j\left(\chi^l \right)H_0^{(1)}\left(k|\mathbf{r}-\mathbf{r^j}| \right)
\label{s2e2}
\end{equation}

\noindent where $\mathcal{D}=\mbox{i}T_0T_0^{'}\left(k R\right) ^2\left\{\left(J_0(k \eta R) \right)^2+\left(J_1(k \eta R) \right)^2 \right\}/4$ and $\zeta^l=\left(\left(\eta^l\right)^2-\eta^2\right)$. $\mathcal{D}$ is found to be independent from both the location and the material characteristics of the defect, while $\zeta$ is the contrast function. Localization and characterization of the defect can thus be fully decoupled.

Let us introduce the normalized vector $\mathbf{v}=\mathbf{V}/\mathbf{V}\cdot\mathbf{V}^*$ ($\mathbf{V}^{*}$ being the complex conjugate of $\mathbf{V}$) associated with vector $\mathbf{V}=\left[\left(E^{l}(\mathbf{r})-E(\mathbf{r})\right)/\mathcal{D}\right]$ and $N$ normalized vectors $\mathbf{g^j}=\mathbf{G^j}/\mathbf{G^j}\cdot\mathbf{G^j}^*$ associated with vector $ \mathbf{G^j}=\left[\mathcal{A}_0^j \sum_{p=1}^N b_{0}^p\left(\chi^j \right)H_0^{(1)}\left(k|\mathbf{r}-\mathbf{r^p}| \right)\right]$.  The defect is localized whenever $\mathcal{P}^j=|1/(1-\|z^j\|)|$, letting $z^j=\mathbf{g^j}\cdot\mathbf{v}^{*}/\mathbf{g^j}\cdot\mathbf{g^j}^{*}$, is maximum. The function $\mathcal{P}^j$ is derived from \cite{iakovleva}. The parameter $z^j$ corresponds with focalization at the defect location. 

Finally, the optical  index of the $l$-th cylinder can be retrieved (this works quite well as seen next, yet it remains heuristic) by averaging the value of $\left(V(\mathbf{r})/G^l(\mathbf{r})+\eta^2\right)^{\frac{1}{2}}$ over the measured data, i.e. $ \widetilde{\eta}^{l}=\mbox{mean}\left(\left(V(\mathbf{r})/G^l(\mathbf{r})+\eta^2\right)^{\frac{1}{2}}\right)$, wherein $\mbox{mean}\left(x(\mathbf{r})\right)$ means the average value of $x(\mathbf{r})$ over the measured data.

\subsection{Localization and characterisation of two identical defects}

The same assumptions and procedure  as above is followed for two identical defects, which corresponds to the so-denoted $(i,j$)-th DFSC configuration. The first-order Born approximation now requires that the field at the location of one defect be only weakly modified by this defect but also by the other one. This means that the defects are ``well separated'', i.e., they are not interacting together. The approximation of the integral equation becomes

\begin{equation}
\displaystyle E^{(i,j)}(\mathbf{r})-E(\mathbf{r})\approx \mathcal{D}\zeta \sum_{l=1}^N\left(\mathcal{A}_0^i b_{0}^l\left(\chi^i \right)+\mathcal{A}_0^j b_{0}^l\left(\chi^j \right)\right)H_0^{(1)}\left(k|\mathbf{r}-\mathbf{r^l}| \right)
\label{s3e1}
\end{equation}

Similarly with what has been done in the previous subsection, one is introducing the normalized vector $\mathbf{v}$ associated with vector $\mathbf{V}=\left[\left(E^{(i,j)}(\mathbf{r})-E(\mathbf{r})\right)/\mathcal{D}\right]$ and $N \times N$ normalized vectors $\mathbf{g^{(q,l)}}$ associated with vector $ \mathbf{G^{(q,l)}}=\left[ \sum_{p=1}^N \left(\mathcal{A}_0^q b_{0}^p\left(\chi^q\right)+\mathcal{A}_0^l b_{0}^p\left(\chi^l\right) \right)H_0^{(1)}\left(k|\mathbf{r}-\mathbf{r^p}| \right)\right]$.  The defect is now localized whenever $\mathcal{P}^{(q,l)}=|1/(1-\|z^{(q,l)}\|)|$, letting $z^{(q,l)}=\mathbf{g^{(q,l)}}\cdot\mathbf{v}^{*}/\mathbf{g^{(q,l)}}\cdot\mathbf{g^{(q,l)}}^{*}$, is maximum.
The optical  index of the $(i,j)$-th cylinders then follows as $ \widetilde{\eta}^{(i,j)}=\mbox{mean}\left(\left(V(\mathbf{r})/G^{(i,j)}(\mathbf{r})+\eta^2\right)^{\frac{1}{2}}\right)$.

\section{Numerical results}

Data are computed by use of the multipole method. The infinite sums $ \sum_{m=-\infty}^{\infty}$ are truncated to $\sum_{m=-M}^{M}$ such that $M=\mbox{int}\left((k\eta R)^{\frac{1}{3}}+k\eta R +5 \right)$, where $\mbox{int}(x)$ is the entire part of $x$. For the inverse problem, one does not make use of the fact that the scattered field is isotropic for the calculation of the specific Green's function as it is proposed in \cite{Felbacq} in the low-frequency approximation. The latter is calculated directly, by multiplying the vector $\mathbf{b}$ by a matrix $\mathbf{H}=[H_0^{(1)}\left(k|\mathbf{r}-\mathbf{r}^j| \right)]$, which is stored once. In the same fashion, the available asymptotic formulae of the reflection and transmission coefficients are not employed; the formulae of these coefficients are as in \cite{Asatryan}.

The wavelength and the optical index $\eta$ are set to $\lambda=20$ and $\eta=2.9$. The low-frequency approximation is then valid for $R$ small enough to have $k\eta R<\!<1$. In the following one assumes that $R=0.15$ and one mostly investigates the localization and characterization of defects located in the central part of the PCs when $\pi/2$.

\begin{figure}[H]
\centering{\includegraphics[width=8.0cm]{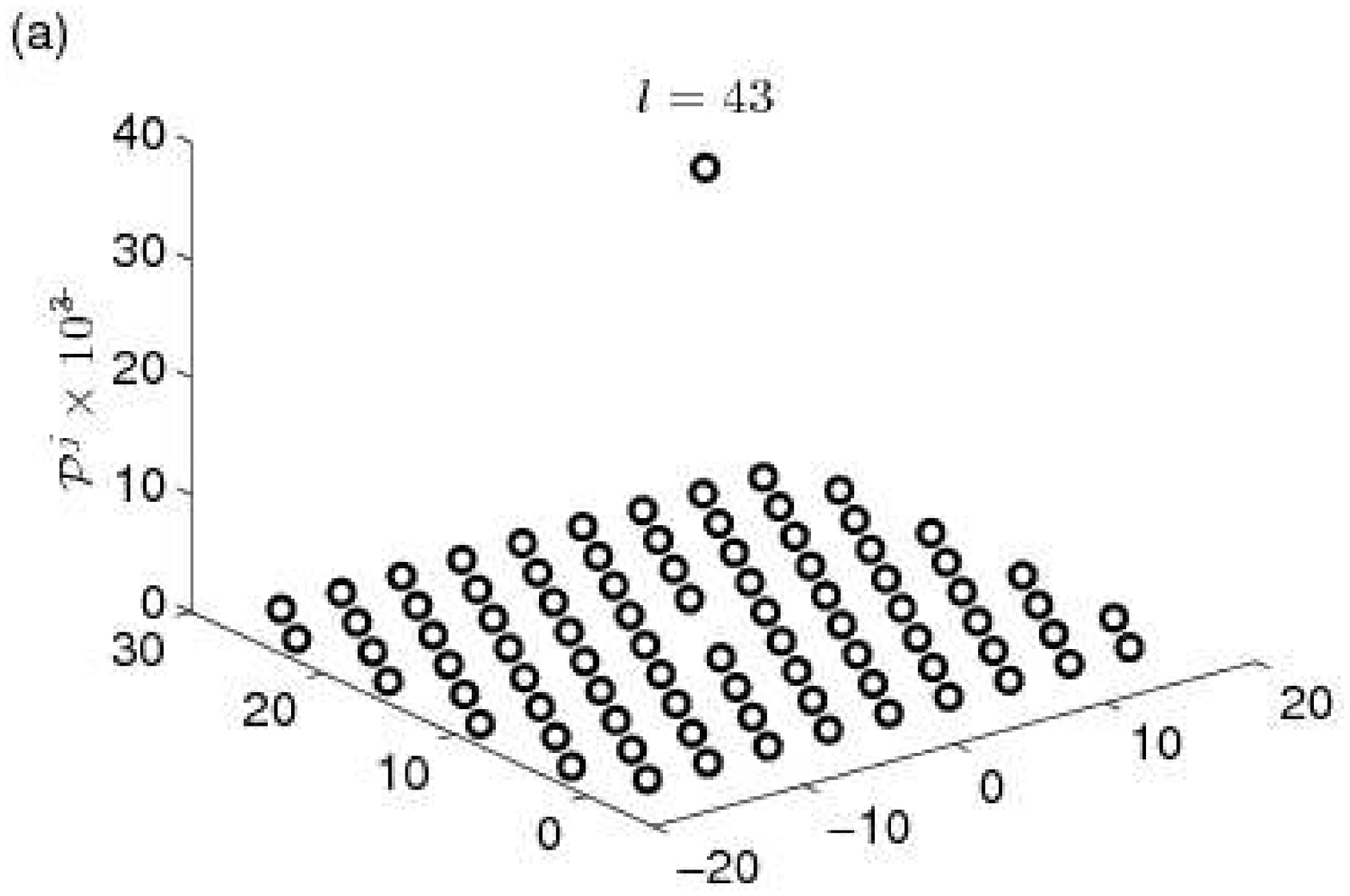}
\includegraphics[width=8.0cm]{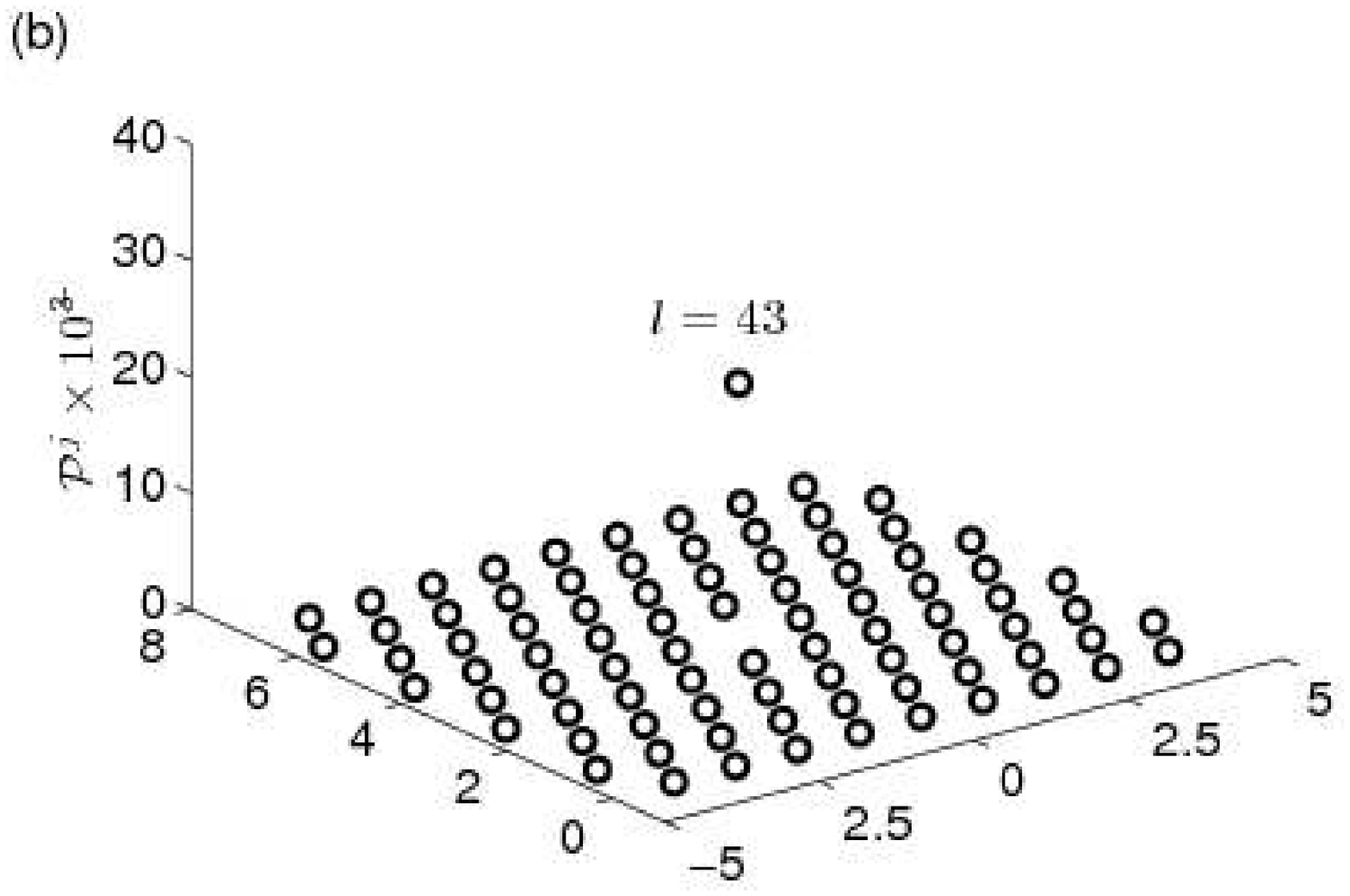}
}
\caption{Crystal with one defect: $R=0.15$, $\eta^{43}=1$ and $\theta^{inc}=\pi/2$. (a) $\mathcal{P}^j \times 10^{-3}$ when $d=4$ ($\widetilde{\eta}^{43}=1.1$) and (b) $\mathcal{P}^j \times 10^{-3}$ when $d=1$ ($\widetilde{\eta}^{43}=1.08$).}
\label{fig1}
\end{figure}

Figure \ref{fig1} shows $\mathcal{P}^j$ for a $43$-th DFSC, with $\eta^l=1$ (i.e., the $43$-th cylinder is removed) and $\theta^{inc}=\pi/2$ when $d=4$ ($R/d=3.75\times10^{-2}$) and $d=1$ ($R/d=0.15$). In both cases, the modified cylinder is clearly retrieved and $\eta^l$ is found with a relative error on its real part $ Er=\left(\Re(\widetilde{\eta}^l)-\eta^l\right)/\eta^l$ less than $0.1$. The $43$-th cylinder seems (in the sense that the function $\mathcal{P}^l$ does not point to another cylinder) to be retrieved with an accuracy of $d$ that is much smaller than the Rayleigh criterion $\lambda/2=10$. 

\begin{figure}[H]
\centering{
\includegraphics[width=8.0cm]{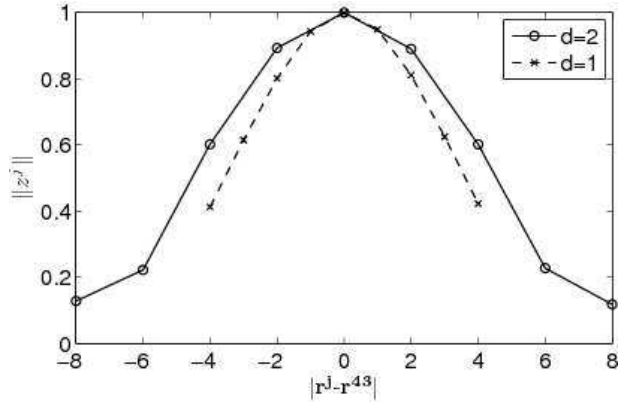}
}
\caption{Crystal with a single defect: $R=0.15$, $\eta^{43}=1$ and $\theta^{inc}=\pi/2$. Cut of $\|z^l\|$ along the axis going through $\mathbf{r^{43}}$ with an angle $\theta=\pi/6$.}
\label{fig2}
\end{figure}

However, this is partly due to the representation chosen, i.e., the value of $\mathcal{P}^l$ is very large since $\|z^l\|\approx 1$, focusing beyond the Rayleigh criterion playing its part only to some extent. Indeed,  strictly speaking, super-resolution would mean that the width at half-height of $\|z^j\|$ is (significantly) smaller than half a wavelength. A cut of $\|z^j\|$ along the axis passing through $\mathbf{r^{43}}$ with an angle of $\pi/6$ is displayed in Fig. \ref{fig2} for $\theta^{inc}=\pi/2$, when $d=2$ ($R/d=7.5\times 10^{-2}$) and $d=1$. Super-resolution, in terms of focusing accuracy, is validated in case $d=1$ and not so much in case $d=2$.

Let us notice that the problem  at hand can be interpreted as the retrieval of the location of an induced line source within a cylinder, whilst the problem attacked in \cite{fink} consists in the retrieval of a line source located outside all cylinders. In this sense, the problem here is quite different but like effects are observed.

\begin{figure}[H]
\centering{\includegraphics[width=8.0cm]{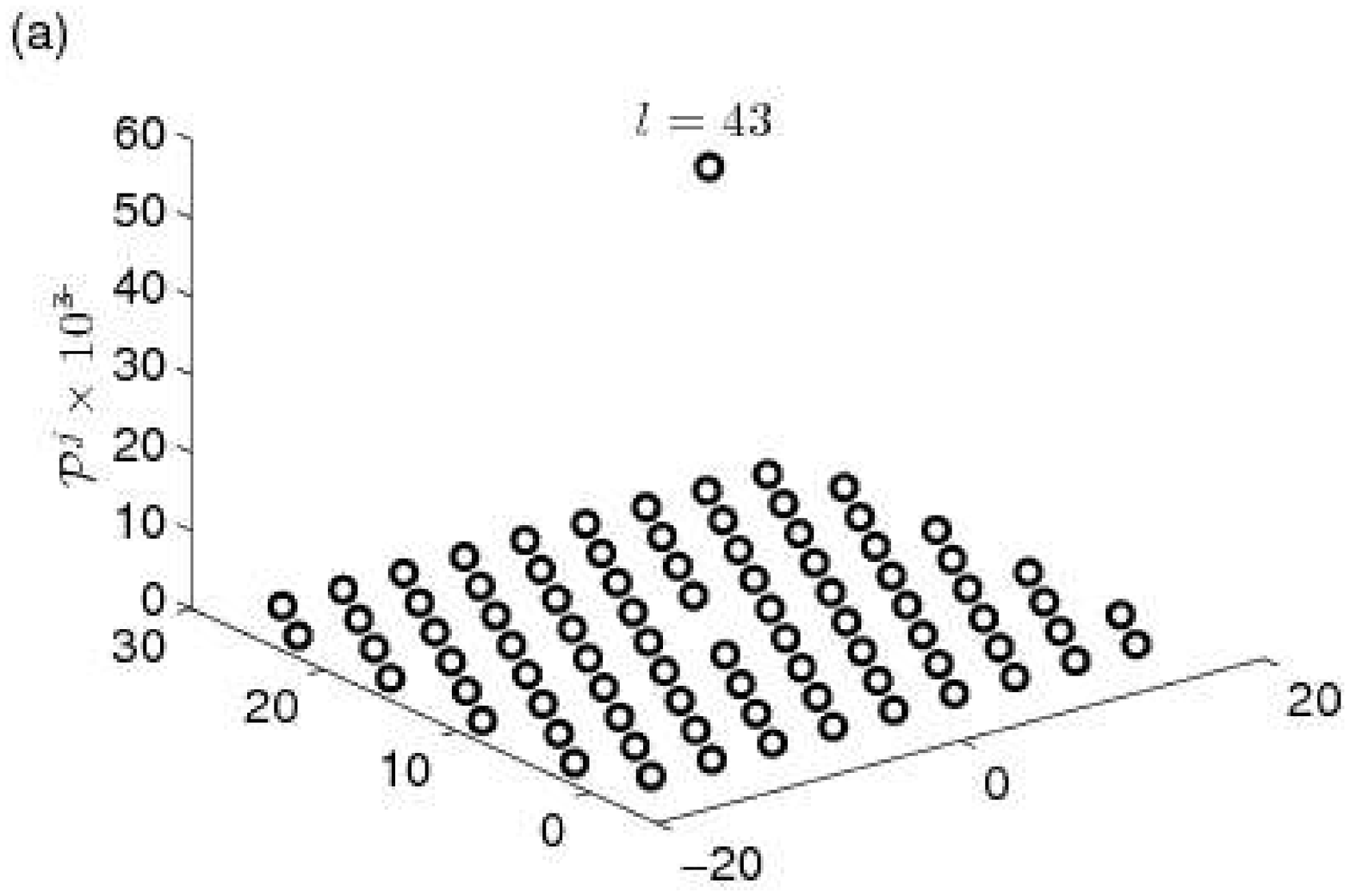}
\includegraphics[width=8.0cm]{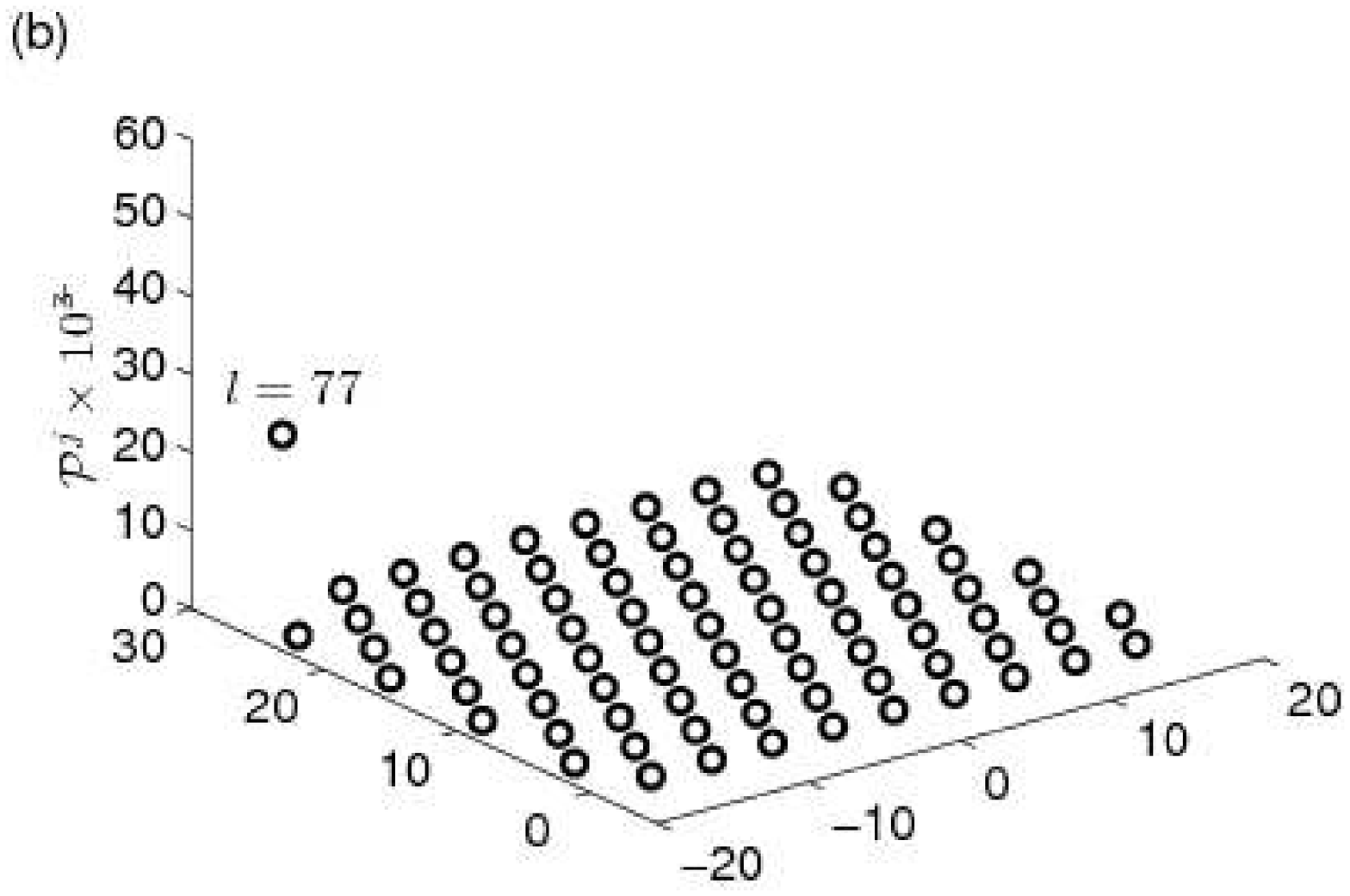}
}
\caption{Crystal with a single defect when $d=4$: $R=0.15$. (a) $\mathcal{P}^j \times 10^{-3}$ when $\eta^{43}=1$ and $\theta^{inc}=\pi/4$ ($\widetilde{\eta}^{43}=1.1$) and (b) $\mathcal{P}^j \times 10^{-3}$  when $\eta^{77}=1$ and $\theta^{inc}=\pi/2$ ($\widetilde{\eta}^{77}=1.1$).}
\label{fig3}
\end{figure}

The retrieved value of $\widetilde{\eta}^l$ when $d=4$ is the same when $\theta^{inc}$ is varied within $[0;\pi/2]$ as well as for a defect which is not located in the central part of the PCs, whilst the height of the peak of $\mathcal{P}^l$ depends on both $\theta^{inc}$ and defect location, Fig. \ref{fig3}, with no obvious rule however.

\begin{figure}[H]
\centering{\includegraphics[width=8.0cm]{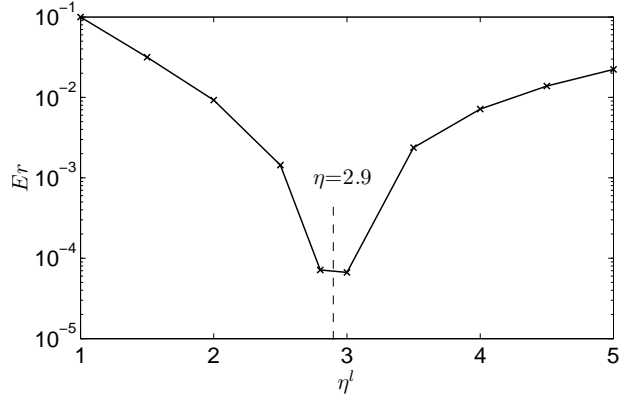}}
\caption{Crystal with a single defect when $d=4$: $R=0.15$ and $\theta^{inc}=\pi/2$. Relative error on the reconstructed value $\widetilde{\eta}^{43}$ for various $\eta^{43}$. The modified cylinder is systematically retrieved.}
\label{fig4}
\end{figure}

Figure \ref{fig4} shows the relative error on the reconstruction of $\widetilde{\eta}^{43}$ for various values of $\eta^{43}$, with $\theta^{inc}=\pi/2$, when $d=4$. The smaller the contrast $\zeta^l$ is, the better the retrieval of $\widetilde{\eta}^{l}$ is. In particular, one is able to retrieve a $\eta^{43}=2.8$, that represents a variation of $3.45\%$ from $\eta$, with a relative error less than $10^{-4}$. The results remain accurate when the low-frequency approximation is not anymore valid at the defect location yet remains valid for cylinders that are constituting the structured background.
\begin{figure}[H]
\centering{\includegraphics[width=8.0cm]{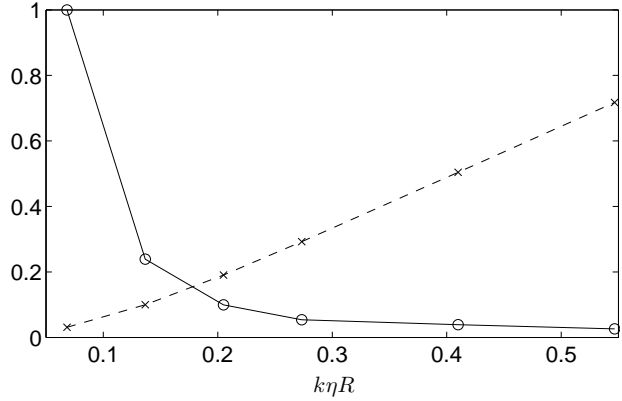}}
\caption{Crystal with a single defect when $d=4$: $\eta^l=1$ and $\theta^{inc}=\pi/2$. Evolution of $Er$ (dashed line) and of the height of the peak of $\mathcal{P}^l$ normalized by its value for $R=0.075$ (solid line) for various values of $R$.}
\label{fig5}
\end{figure}

The relative height $h^l$ of the peak of $\mathcal{P}^l$ is defined by $h^l=\mathcal{P}^l-\mbox{mean}_{j\neq l}\left(\mathcal{P}^j\right)$, wherein $\mbox{mean}_{j}\left(x^j\right)$ is the average value of $x^j$ over $j\in[1,N]$. The evolution of $Er$ and of $h^l$ with $R\in[7.5\times 10^{-2},0.6]$ is shown in Fig. \ref{fig5} with $\theta^{inc}=\pi/2$ when $d=4$ and $\eta^l=1$. It is observed that, even though the low-frequency approximation does not hold for the background, the modified cylinder is retrieved, but the reconstructed values $\widetilde{\eta}^l$ are not accurate anymore. When $\eta$ is varied, the reconstructed values (in particular the imaginary part) of $\eta^l$ are now inaccurate even for $k\eta R=0.27$. This means that the result, and as a matter of fact the use of the first-order Born approximation, remains much more appropriate when $R$ becomes large than when $\eta$ becomes large.

\begin{figure}[H]
\centering{\includegraphics[width=8.0cm]{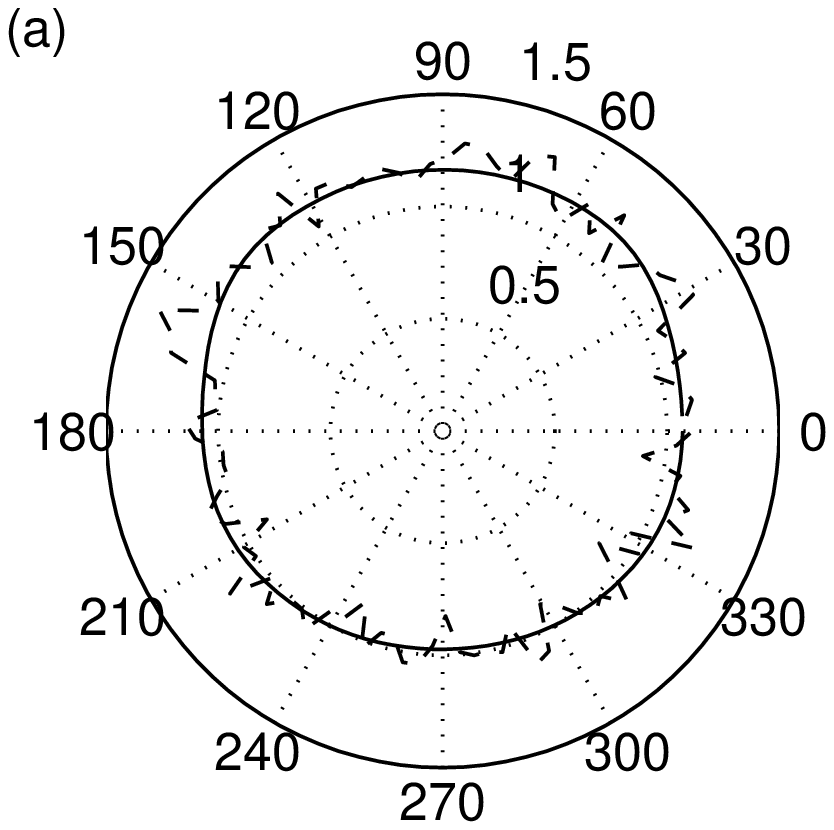}
\includegraphics[width=8.0cm]{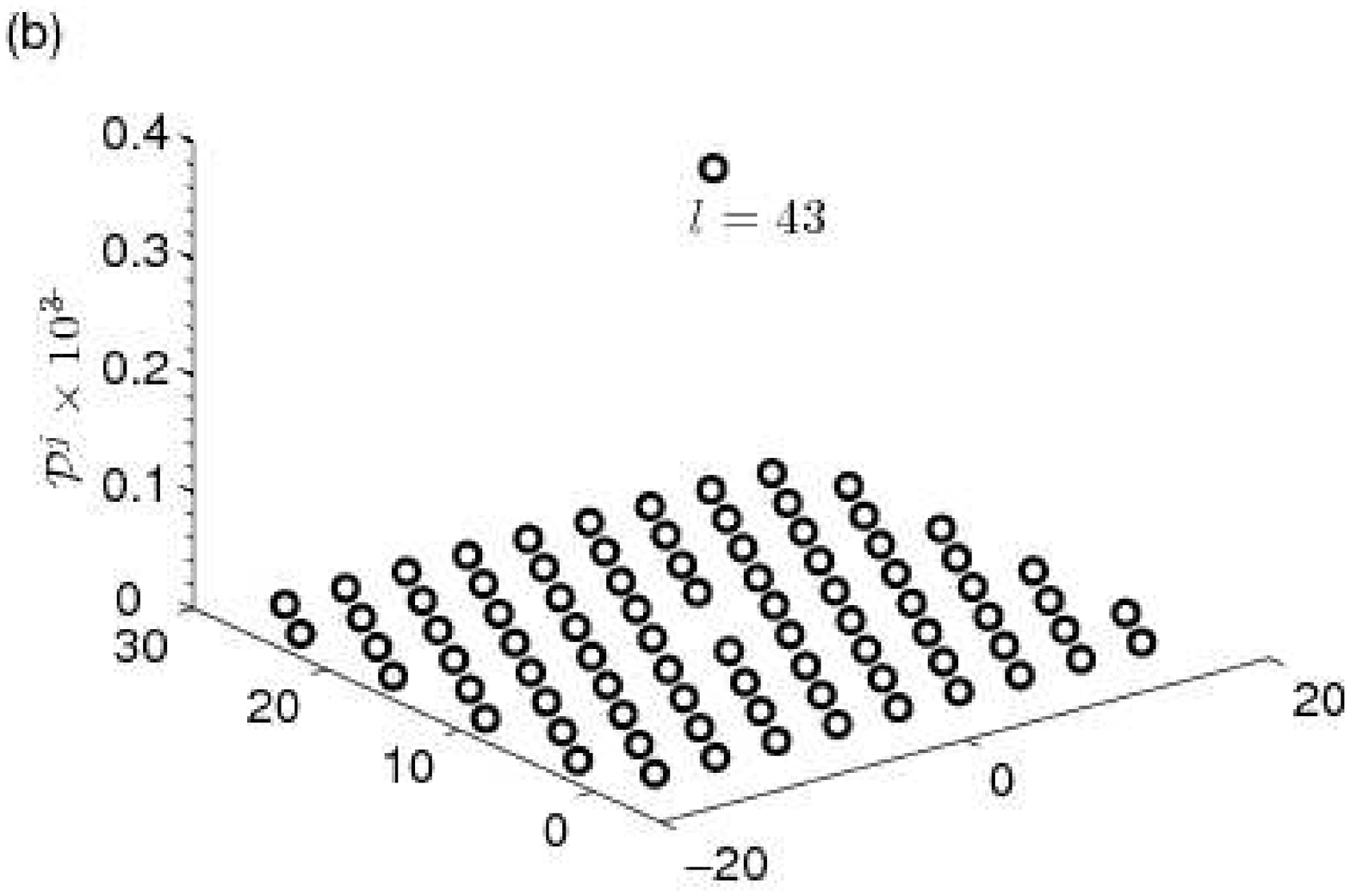}
}
\caption{Crystal with a single defect when $d=4$: $R=0.15$, $\eta^l=1$ and $\theta^{inc}=\pi/2$. (a) the solid curve depicts $\|\mathbf{v}\|$ and the dashed curve depicts $\|\mathbf{v}_{noise}\|$, and (b) $\mathcal{P}^l \times 10^{-3}$ ($\widetilde{\eta}^{43}=1.1$).}
\label{fig6}
\end{figure}

When a white Gaussian noise ($SNR=50dB$) is added to both real and imaginary part of $E$ and $E^l$, the corresponding normalized vector is denoted by $\mathbf{v}_{noise}$. Both $\|\mathbf{v}_{noise}\|$ and $\|\mathbf{v}\|$ are plotted in Fig. \ref{fig6}. The missing cylinder is imaged and $\widetilde{\eta}^l$ is retrieved with a relative error less than $0.1$. The main impact of the addition of a white Gaussian noise to the data is a decrease of $h^l$.

\begin{figure}[H]
\centering{\includegraphics[width=8.0cm]{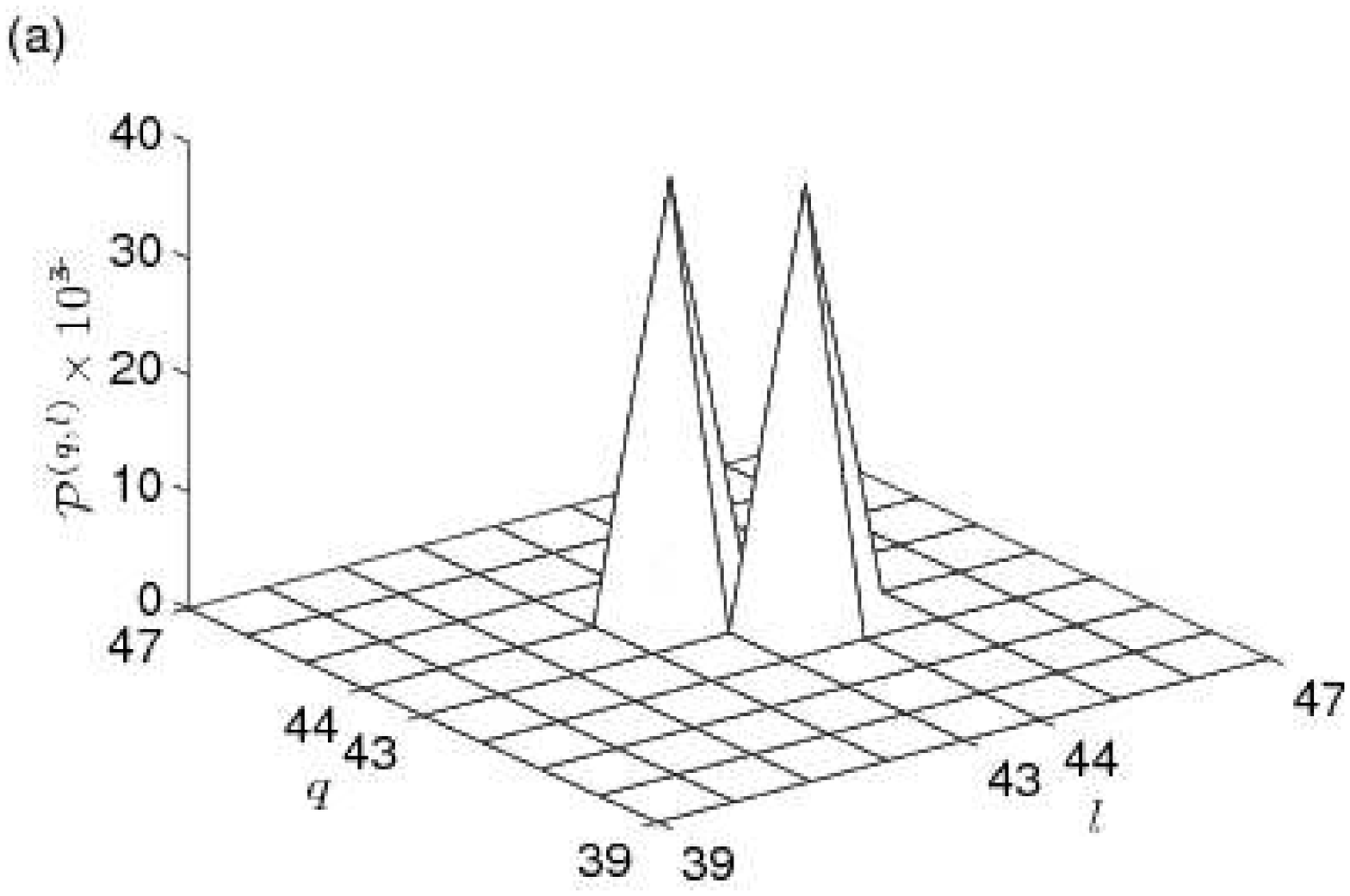}
\includegraphics[width=8.0cm]{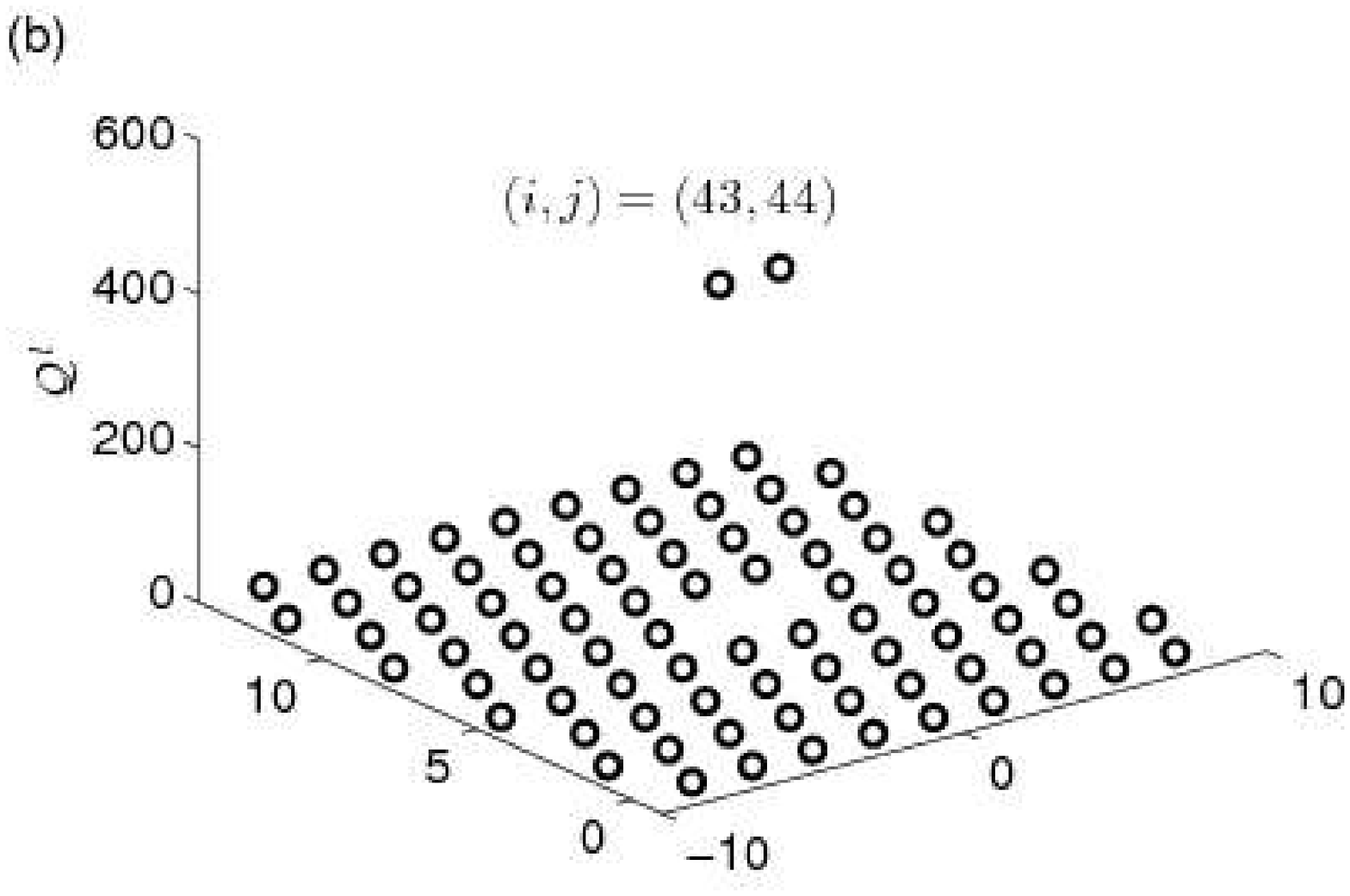}}
\caption{Crystal with two defects close to one another when $d=2$: $R=0.15$, $\eta^l=1$ and $\theta^{inc}=\pi/2$. (a) $\mathcal{P}^{(q,l)} \times 10^{-3}$ ($\widetilde{\eta}^l=1.1$) and (b) $\mathcal{Q}^{l}$.}
\label{fig7}
\end{figure}

As for the $(i,j)$-th DFSC configuration, the major difficulty is in the recovery of modified cylinders close to one another. $\mathcal{P}^{(q,l)}$ is symmetric in terms of $q$ and $l$. Let us define the vector $\mathcal{Q}^{q}$ such that $\mathcal{Q}^{q}=\mathcal{Q}^{l}=\mbox{mean}_l\left(\mathcal{P}^{(q,l)}\right)$, which enables us to depict the results in the same form as in the case of the $(l)$-th DFSC configuration. Figure \ref{fig7} shows both $\mathcal{P}^{(q,l)}$ and $\mathcal{Q}^{q}$ for a $(43,44)$-th DFSC configuration with $\theta=\pi/2$ when $d=2$. Two identically modified cylinders are retrieved, separated from a distance $d$ which is smaller than $\lambda/2$.

\section{Conclusion}

The simple tomographic algorithm described herein appears as a first step in the monitoring of structured materials like PCs. Use of the specific Green's function together with the first-order Born approximation enables us to localize and characterize simple defects consisting in absence of cylinders or identical modification of the optical index of cylinders in finite-size PCs. Several results exhibit a possible retrieval of two defects beyond the Rayleigh criterion. The algorithm can be used as a first iteration in an iterative solution scheme.  

Various difficulties would arise if the low-frequency approximation was no more valid, which is the case for usual PCs at the location of band gap (their frequency band of use). The dependance of $h^l$ on the angle of incidence of the plane wave sollicitation as well as its dependance on the defect location is still a challenge.

\end{document}